\documentclass[amsmat,amssymb,amsfonts,aps,prb,twocolumn,superscriptaddress]{revtex4-2}

\usepackage{graphicx}
\usepackage{xcolor}

\newcommand{\bra}[1]{\left\langle#1\right|}
\newcommand{\ket}[1]{\left|#1\right\rangle}

\newcommand{\Tr}{{\rm Tr}}
\renewcommand{\Im}{{\rm Im}}
\renewcommand{\Re}{{\rm Re}}

\newcommand{\CG}{{\cal G}}

\newcommand{\Br}{{\bf r}}

\newcommand{\BA}{{\bf A}}

\newcommand{\BR}{{\bf R}}

\begin{document}

\title{Renormalization of spin excitations and Kondo effect in open shell nanographenes}

\author{David Jacob}
\email{david.jacob@ehu.es}
\affiliation{Departamento de Pol\'{i}meros y Materiales Avanzados: F\'{i}sica, Qu\'{i}mica y Tecnolog\'{i}a, Universidad del Pa\'{i}s Vasco UPV/EHU,
  Av. Tolosa 72, E-20018 San Sebasti\'{a}n, Spain}
\affiliation{IKERBASQUE, Basque Foundation for Science, Plaza Euskadi 5, E-48009 Bilbao, Spain}

\author{Ricardo Ortiz}
\affiliation{Departamento de F\'{i}sica Aplicada, Universidad de Alicante, E-03690 San Vicente del Raspeig, Spain}
\affiliation{Departamento de Qu\'{i}mica F\'{i}sica, Universidad de Alicante, E-03690 San Vicente del Raspeig, Spain}

\author{Joaqu\'{i}n Fern\'{a}ndez-Rossier}
\affiliation{QuantaLab, International Iberian Nanotechnology Laboratory (INL), 4715-330 Braga, Portugal}
\altaffiliation[On leave from ]{Departamento de F\'{i}sica Aplicada, Universidad de Alicante, E-03690 San Vicente del Raspeig, Spain}
\date{\today}

\begin{abstract}
  We study spin excitations and Kondo effect in open-shell nanographenes, motivated
  by recent scanning tunneling inelastic spectroscopy experiments. Specifically, we
  consider three systems, the triangulene, the extended triangulene with rocket shape,
  both with an $S=1$ ground state, and a triangulene dimer with $S=0$ on account of
  intermolecular exchange. We focus on the consequences of hybridization of the nanographene
  zero-modes with a conducting substrate on the $dI/dV$ lineshapes associated with spin excitations.
  The partially filled nanographene zero-modes coupled to the conduction electrons in the substrate
  constitute multi-orbital Anderson impurity models that we solve in the one-crossing approximation
  which treats the coupling to the substrate to infinite order.
  We find that the coupling to the substrate leads to (i) renormalization of the spin 
  flip excitation energies of the bare molecule, (ii) broadening of the spectral
  features and (iii) the emergence of zero bias Kondo peaks for the $S=1$ ground states.
  The calculated substrate induced shift of  the spin excitation energies
  is found to be significantly larger than their broadening, which implies that this effect has to be considered
  when comparing experimental results and theory.
\end{abstract}

\maketitle

\section{Introduction}

Open shell nanographenes (NGs) have been studied theoretically for many
decades~\cite{LonguetHiggings,clar72,Borden77,Fernandez2007,wang2009,morita11,melle15,Ortiz:NL:2019},
on account of their very peculiar magnetic properties, with local moments associated to $\pi$ molecular states, with very 
small magnetic anisotropy and strong quantum spin fluctuations.  
However, their large chemical reactivity was a great obstacle for their experimental study. This situation 
has changed dramatically with the advent of on-surface synthesis techniques~\cite{narita2015,ruffieux2016,Song:CSR:2020}
combined with surface scanning probes techniques, such as atomic force and scanning 
tunneling microscopes (AFM and STM). These techniques make it now  possible to synthetize an increasing number 
of open-shell NGs, such as ribbons with zigzag edges~\cite{wang2016}, the Clar's goblet~\cite{mishra19b}, rhombenes~\cite{mishra2020b},
triangulenes~\cite{pavlivcek2017,su19,mishra2020,Mishra:2021}, heptauthrene~\cite{heptauthrenesynth} and
others~\cite{NachoNat,li2020,zheng2020,mishra19tri,Sanchez-Grande:JPCL:2021,Hieulle:2021},
and to probe their electronic properties with atomic scale resolution.

A prominent challenge in this new research area is to probe the spin properties of open shell NGs~\cite{Ortiz:PSS:2020}.
STM permits to carry out inelastic electron tunnel spectroscopy (IETS) with atomic resolution (for a review 
of this topic see e.g. Ref.~\onlinecite{Reed:materialstoday:2008}). Two types of features signal the presence of open-shell configurations: zero bias Kondo peaks and
the observation of step-like features in the $dI/dV$ spectra at bias voltages $V=\pm V_{\rm ex}$. However, whereas inelastic steps indicate the existence of 
an excitation at energy $eV_{\rm ex}$, their spin dependent origin can only be confirmed directly if application of a magnetic field shifts the energy of that 
excitation.
This approach has been thoroughly used to probe spin excitations of individual magnetic
adatoms~\cite{Heinrich:Science:2004,Hirjibehedin:Science:2007,fernandez2009,Zitko:NJP:2010,Ternes:NJP:2015},
single magnetic molecules~\cite{Tsukahara:PRL:2009} and adatom chains~\cite{hirjibehedin2006spin,Spinelli:NMater:2014,choi2019} in the last two decades.

The direct experimental confirmation of the spin nature of the excitations cannot be carried out when the 
Zeeman shift of the excitation energy, in the range of $g\mu_B\simeq0.12$~meV per Tesla, is smaller than the spectral 
resolution of IETS, controlled by temperature and broadening due to the lifetime of the spin excitations. 
The latter is governed by the strength of the effective exchange interactions to the substrate. In the case 
of NGs deposited directly on gold~\cite{NachoNat,mishra19b,li2020,mishra2020,zheng2020,mishra2020b}, broadening is definitely larger than Zeeman splitting, 
making it necessary to rely on the comparison with theory.
The situation may be different in the case of Kondo effect in $S=1$ NGs where the narrow Kondo peak
splits when the Zeeman energy exceeds the Kondo temperature.
Moreover, as reported recently by Li and coworkers~\cite{li2020}, in the case of an \emph{underscreened}
Kondo effect, due to the residual magnetic moment of the NG, the Zeeman splitting of the Kondo resonance
can already be observed at relatively small magnetic fields, limited only by thermal broadening.

Direct comparison between the excitation energies obtained from high level quantum chemistry calculations and the singlet-triplet excitation energies yield a fairly 
good agreement with the experimental observations in bowtie~\cite{mishra19b,Ortiz:NL:2019} and rhombenes~\cite{mishra2020b}. However, it is known that the coupling
of a magnetic system to the conduction electrons of the substrate not only broadens the spectral function of the 
spin excitations, but it also produces a renormalization of the energies, i.e., a shift of the step
position~\cite{Korytar:PRB:2012,Oberg:NNano:2014,Delgado:SS:2014,Jacob:EPJB:2016,Jacob:PRB:2018},
which has to be taken into account when comparing theory and experiment.
In the case of a (spin) degenerate ground state (GS), the coupling to the substrate gives rise to the Kondo effect,
screening the magnetic moment of the adsorbate~\cite{Hewson:book:1997}.
This certainly occurs in NGs whose GS has $S=1/2$, but it has also been observed in NGs 
predicted to have $S=1$~\cite{li2020,heptauthrenesynth}.
Note that the NG hosts the localized electrons, while the conduction electrons are in the substrate.
This is different from the case of Kondo effect of magnetic impurities and point defects in
graphene~\cite{Sengupta:PRB:2008,Cornaglia:PRL:2009,Wehling:PRB:2010,Jacob:PRB:2010a,Fritz_2013,Principi:PRB:2015,Shi:PRB:2019}.

In this work we address all three effects, i.e., broadening and shift of IETS steps and the emergence
of zero bias peaks, in the same theoretical framework. To the best of our knowledge, previous theory work has not addressed these
  in the context of nanographenes. We model the NGs using the Hubbard Hamiltonian
solved by means of exact numerical diagonalization in the subspace defined 
by the zero-modes (ZMs) of the NG. The energies obtained for this model compare well with higher 
level quantum chemistry calculations~\cite{Ortiz:NL:2019,mishra2020b}. The NG coupled to the 
substrate is then described in terms of an Anderson impurity model (AIM). Here we make no 
attempt to describe the substrate from first principles. Instead, we use the strength of the coupling
to the substrate, assumed to be the same for all atoms in the NG, as a tunable parameter in
the calculation. The AIM is solved in the one-crossing approximation (OCA) which consists in a 
diagrammatic expansion in the coupling strength to infinite order~\cite{Haule:PRB:2001}. 
This yields the spectral function of the ZMs of the NG flake, directly related to 
the $dI/dV$ in the tunneling regime~\cite{Jacob:PRB:2018,Jacob:NL:2018}.

\begin{figure}
  \includegraphics[width=\linewidth]{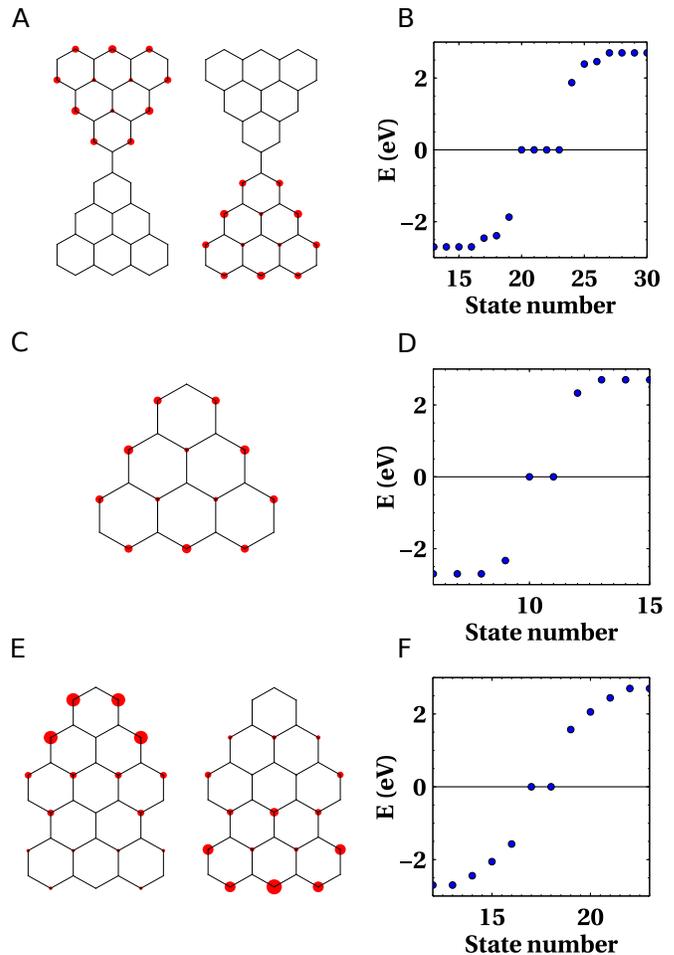}
  \caption{\label{fig:structures}
    Atomic structures with graphical representations of the ZMs (A,C,E) and corresponding single-particle 
    spectra (B,D,F) for $t=-2.7$~eV of the three NGs discussed in the paper: triangulene dimer (A,B), 
    single triangulene (C,D), and 'rocket' structure (E,F). 
    The size of the red circles in the structures (A,C,E) yield the weight 
    $|\psi_\alpha(i)|^2$ of the $E=0$ wavefunction $\psi_\alpha$ on carbon site $i$.
    Note that the other two ZMs in case of the dimer and the other ZM for the
    single triangulene have the same weights (but different phases) as the ones already depicted
    in A and C.
  }
\end{figure}

We choose three NG systems that have been studied experimentally, shown in Fig.~\ref{fig:structures}: 
(i) antiferromagnetically coupled dimers~\cite{mishra2020} of $S=1$ triangulenes with an $S=0$ GS, 
(ii) $S=1$ graphene triangulene~\cite{pavlivcek2017}, and (iii) $S=1$ extended triangulene~\cite{li2020},
refered to as 'rocket' structure from now on.

\section{Model and Method}

We consider the following Hamiltonian to descibe a NG flake on a metallic substrate:
\begin{equation}
  \label{eq:model}
  {\cal H}= {\cal H}_{\rm ng}+ {\cal H}_{\rm sub} + {\cal V}_{\rm hyb}.
\end{equation}

The first term in (\ref{eq:model}) describes the NG flake in terms of a Hubbard model~\cite{Ortiz:NL:2019}:
\begin{equation}
  \label{eq:NG}
  {\cal H}_{\rm ng} = \sum_{\langle i,j \rangle\atop \sigma} t \,(c_{i\sigma}^\dagger c_{j\sigma} + c_{j\sigma}^\dagger c_{i\sigma}) 
  +  \sum_i ( \epsilon_0 \, n_{i} + U \,n_{i\uparrow} n_{i\downarrow} )
\end{equation}
where $c_{i\sigma}^\dagger$ ($c_{j\sigma}$) create (destroy) an electron of spin $\sigma\in\{\uparrow,\downarrow\}$ 
at carbon site $i$ ($j$) of the NG, $n_{i\sigma}=c_{i\sigma}^\dagger c_{i\sigma}$ and $n_i=\sum_\sigma n_{i\sigma}$.
The model is determined by the first neighbour hopping $t$ (second and third  
neighbour hoppings are neglected here), the atomic Hubbard $U$, and the carbon 
onsite energy $\epsilon_0$. We assume $t=-2.7$~eV and take $U$ as an adjustable 
parameter in the range of  $U\simeq1.5|t|\simeq4$~eV. First principles estimates yield  $U\simeq9$~eV for graphene~\cite{wehling2011}.
The fact that smaller values provide a better agreement with the experiments is probably due to screening of the Coulomb interaction by the conducting substrate.

The second term in (\ref{eq:model}) describes the conduction electron bath in the 
substrate, ${\cal H}_{\rm sub} = \sum_{k,\sigma} \epsilon_k \, b_{k\sigma}^\dagger b_{k\sigma}$,
where $b_{k\sigma}$ ($b_{k\sigma}^\dagger$) destroys (creates) an electron in bath state $k$ with 
spin $\sigma$. 
Without loss of generality we set the chemical potential of the conduction electron 
bath to zero ($\mu=0$). 
The third term in (\ref{eq:model}) is the hybridization between the NG states
and the conduction electron bath, 
${\cal V}_{\rm hyb} = \sum_{k,\sigma,i} V_{k,i} \, b_{k\sigma}^\dagger c_{i\sigma} + {\rm h.c.}$
Integrating out the conduction electron degrees of freedom, we obtain 
the hybridization function, $\Delta_i(\omega) = \sum_k |V_{k,i}|^2/(\omega^+-\epsilon_k)$,
which describes shift (real part) and broadening (imaginary part) of carbon site $i$ due to
the coupling to the bath; as usual $\omega^+=\omega+i\eta$ shifting the poles/branch cut
of the bath Greens function $1/(\omega-\epsilon_k)$ from the real axis infinitesimally to the lower complex plain in order to make
$\Delta_i(\omega)$ analytical in the upper complex plain.
Here we assume the wide-band limit, i.e. the single-particle broadening $\Gamma_i\equiv-\Im\,\Delta_i(\omega)$ 
is assumed to be constant within a large energy window, and thus the real part of the hybridization function vanishes, $\Re\,\Delta_i(\omega)=0$. 
Moreover, we assume that each carbon site couples equally to the substrate, i.e. $\Gamma_i=\Gamma={\rm const}$.
These are reasonable approximations for noble metal substrates in the low-energy (or low-bias) regime that 
we are interested in. 

As the many-body Hilbert space grows exponentially with the number of single-particle orbitals (or sites), 
it is not possible to numerically diagonalize (\ref{eq:NG}) for the entire NG.
One possibility is to treat the NG in the Dynamical Mean-Field Approximation (DMFA)
adapted to nanoscale systems~\cite{Jacob:PRB:2010b,Valli:NL:2018}.
However, DMFA neglects non-local correlations which are known to be important in low dimensional systems.
Instead we focus on a small subspace C of the full many-body Hilbert space 
defined by a subset of the molecular orbitals $\{\psi_\alpha\}$ of the NG. 
Specifically, we choose the so-called ZMs which are the zero-energy eigenstates of the hopping part of (\ref{eq:NG})
which host the unpaired electrons, responsible for the magnetism 
of open shell NGs~\cite{Fernandez2007,wang2009,Ortiz:NL:2019,Ortiz:PSS:2020}.
The ZMs can be expanded in the site basis $\{\ket{i}\}$  as $\ket{\psi_\alpha}=\sum_i\psi_\alpha(i)\ket{i}$.
Fig.~\ref{fig:structures} shows graphical representations of the ZMs for the three NG systems considered here.

The subspace C coupled to the substrate defines a \emph{multi-orbital} AIM.
The impurity shell is given by the projection of the NG Hamiltonian (\ref{eq:NG}) onto C and takes the form
\begin{equation}
  \label{eq:HC}
  {\cal H}_{\rm C} = \sum_\alpha \varepsilon_{\rm C} \, N_\alpha + \sum_{{\alpha,\alpha^\prime,\beta,\beta^\prime}\atop{\sigma,\sigma^\prime}} 
  \tilde{U}_{\alpha\alpha^\prime\beta\beta^\prime} \, C_{\alpha\sigma}^\dagger C_{\alpha^\prime\sigma^\prime}^\dagger C_{\beta^\prime\sigma^\prime} C_{\beta\sigma}
\end{equation}
where $C_{\alpha\sigma}^\dagger$ ($C_{\alpha\sigma}$) creates (destroys) an electron in molecular orbital $\psi_\alpha$ in subspace C 
with spin $\sigma$, $N_\alpha=\sum_\sigma C_{\alpha\sigma}^\dagger C_{\alpha\sigma}$ and $\tilde{U}_{\alpha\alpha^\prime\beta\beta^\prime}$ is the Coulomb matrix. 
Note that the Hubbard interaction which is local in the site-basis becomes more complicated in the basis of the molecular orbitals, 
$\tilde{U}_{\alpha\alpha^\prime\beta\beta^\prime}=U \sum_i \psi_\alpha^\ast(i) \psi_{\alpha^\prime}^\ast(i) \psi_\beta(i) \psi_{\beta^\prime}(i)$.
On the other hand the coupling to the substrate, described by the hybridization function,
is still the same for all ZMs and constant, i.e., $\Delta_\alpha(\omega)=-i\Gamma$, since 
we have assumed it to be site-independent.
$\varepsilon_{\rm C}$ in (\ref{eq:HC}) is the effective single-particle energy of the ZMs, related to
the on-site energy $\epsilon_0$ of the NG, cf. eq.~(\ref{eq:NG}), but modified by the mean-field interaction of C
with the rest of the NG. $\varepsilon_{\rm C}$ is chosen such that C is close to half-filling, 
i.e. close to the particle-hole (ph) symmetric point corresponding to the neutral isolated NG,
given by $\varepsilon_{\rm C}^\ast=-\tilde{U}/2-(\tilde{U}^\prime-\tilde{J}_{\rm H}/2)(N_{\rm C}-1)$~\cite{Jacob:PRB:2018}
where $\tilde{U}=\tilde{U}_{\alpha\alpha\alpha\alpha}$ is the intra-orbital Coulomb,
$\tilde{U}^\prime=\tilde{U}_{\alpha\beta\alpha\beta}$ the inter-orbital Coulomb and
$\tilde{J}_{\rm H}=\tilde{U}_{\alpha\beta\beta\alpha}$ the Hund interaction
between the ZMs $\alpha\neq\beta$ in C.
Charge transfer from the substrate leads to detuning of $\epsilon_{\rm C}$ from the ph symmetric point.
In our model the detuning from ph symmetry, $\delta\varepsilon_{\rm C}=\varepsilon_{\rm C}-\varepsilon_{\rm C}^\ast$, is taken as an adustable parameter.

We solve the multi-orbital AIM within the One-Crossing Approximation (OCA)~\cite{Haule:PRB:2001,Haule:PRB:2010}.
In short, OCA consists in a diagrammatic expansion of the propagators $G_m(\omega)$ 
associated with the many-body eigenstates (or pseudo particles (PPs)) $\ket{m}$ with eigenenergies $E_m$ 
of the \emph{isolated} impurity Hamiltonian 
(\ref{eq:HC}) in terms of the hybridization function $\Delta_\alpha(\omega)$ to infinite order, 
but summing only a subset of diagrams (only those where conduction electron lines cross at most once).
The PP selfenergy $\Sigma_m(\omega)$ describes the interaction of the PP $m$ with other PPs 
via the conduction electron bath. Its real part yields the renormalization of the
many-body eigenenergies $E_m$ and correspondingly of the electronic excitations due to 
coupling to the bath.
The electronic spectral functions $A_\alpha(\omega)$ for the NG ZMs 
are then obtained from convolutions of the PP propagators $G_m(\omega)$,
see App.~\ref{app:OCA} for more details.

In the ideal STM limit (vanishing coupling to the STM tip) the NG flake is in equilibrium with
the substrate. Making use of the Meir-Wingreen formula~\cite{Meir:PRL:1992} a direct relation 
between the differential conductance from the tip to the sample and the equilibrium spectral 
function matrix $\mathbf{A}(\omega)$ of the NG can be derived~\cite{Jacob:NL:2018,Jacob:JPCM:2018}.
In the zero-temperature limit this yields:
\begin{equation}
  \label{eq:dIdV:ZM}
  \frac{dI}{dV} \sim \Tr\left[ \mathbf{\Gamma}^{\rm tip} \mathbf{A}(eV) \right] = \sum_\alpha \gamma^{\rm tip}_\alpha A_\alpha(eV)
\end{equation}
where $\mathbf{\Gamma}^{\rm tip}$ is the coupling matrix between the NG and the STM tip,
and $\mathbf{A}(\omega)$ is the spectral function matrix of the NG. 
In the last step we have further taken into account that for our model the spectral function matrix is diagonal in the 
molecular orbitals $\psi_\alpha$, i.e., $\mathbf{A}(\omega)=\sum_\alpha \ket{\psi_\alpha} A_\alpha(\omega) \bra{\psi_\alpha}$,
and $\gamma^{\rm tip}_\alpha\equiv\bra{\psi_\alpha}\mathbf{\Gamma}^{\rm tip}\ket{\psi_\alpha}$ is the coupling of orbital $\psi_\alpha$
to the STM tip.

If the STM tip is directly above a carbon site $i$, the coupling will be mainly to this site,
i.e. $\mathbf{\Gamma}^{\rm tip}\approx\gamma_i^{\rm tip}\ket{i}\bra{i}$, so that the 
$dI/dV$ corresponds to the local spectral function $\rho_i\equiv\bra{i}\mathbf{A}\ket{i}$ 
of that carbon site $i$, i.e., $dI/dV \sim \gamma^{\rm tip}_i \rho_i(eV)$
where the local spectral function can be calculated as
\begin{equation}
  \rho_i(\omega)=\sum_{\alpha\in{\rm NG}} |\psi_\alpha(i)|^2 A_\alpha(\omega).
\end{equation}
Note that the orbital index $\alpha$ runs over all orbitals $\psi_\alpha$ of the NG 
and not just over the ZMs.

In an STM experiment, measuring the map $\CG(\BR)\equiv\left.\frac{dI}{dV}\right|_{V_0;\BR}$ of a molecule for different positions $\BR$ of the STM tip
and at fixed voltage $V_0=\omega_0/e$ corresponding to a feature in the $dI/dV$ spectrum, e.g., a Kondo peak, inelastic step, or
HOMO/LUMO resonance, gives an (approximate) picture of the orbital associated with that feature.
This $\CG(\BR)$ map roughly corresponds to the spatially resolved spectral function $\rho(\Br;\omega)\equiv\bra{\Br}\mathbf{A}\ket{\Br}$
evaluated at $\omega=\omega_0$ and $\Br=\BR$, which can be calculated easily within our approach as
\begin{equation}
  \rho(\Br;\omega) = \sum_{i,j} \phi_i^\ast(\Br) \rho_{ij}(\omega) \phi_j(\Br)
\end{equation}
where now also the off-diagonal elements of the spectral function in the site basis, $\rho_{ij}(\omega) \equiv \bra{\phi_i} \BA(\omega) \ket{\phi_j} = \sum_\alpha \psi_\alpha(i) A_\alpha(\omega) \psi_\alpha^\ast(j)$
are required. Here we assume a single Slater type $2p_z$-orbital localized at each carbon site of our tight-binding model using Slater's
original parametrization~\cite{Slater:PR:1930}, i.e. $\phi_i(\Br)\sim z\,e^{-\zeta|\Br-\BR_i|}$ where $\Br=(x,y,z)$, $\zeta=1.625/a_0$ for carbon
where $a_0=0.5292$~\r{A} is the Bohr radius and $\BR_i=(X_i,Y_i,0)$ is the position of carbon site $i$ (assuming the NG is located in the $z=0$ plane). 
The carbon sites of the NG flakes are assumed to be at the positions of a perfect graphene lattice with nearest-neighbor distance $a=1.42$~\r{A}.

\section{Results}

\subsection{Triangulene dimer}

\begin{figure*}[ht]
  \includegraphics[width=\linewidth]{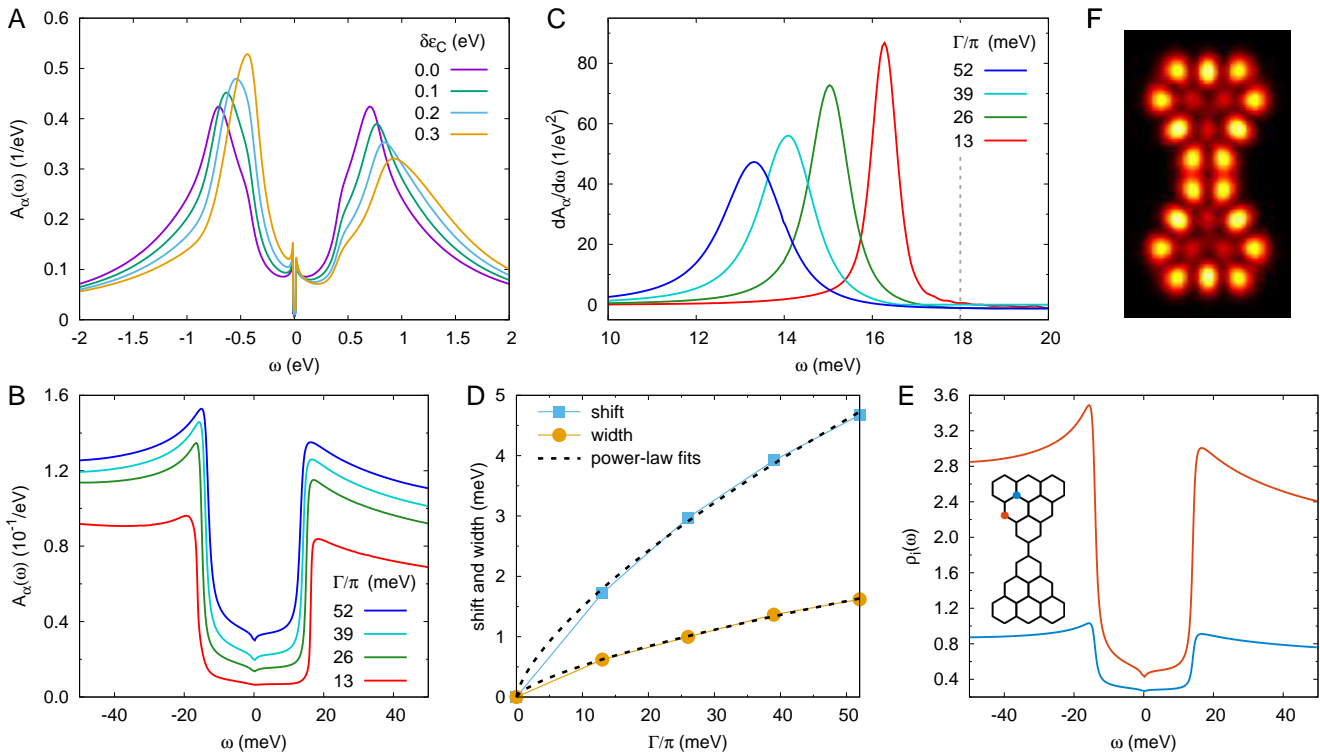}
  \caption{\label{fig:dimer}
    OCA results for triangulene dimer for $t=-2.7$~eV, $U=1.9|t|$ and $\varepsilon_{\rm C}^\ast=-0.485$~eV.
    (A) Spectral funtion of zero modes for different values of the on-site energy shift $\delta\varepsilon_{\rm C}$
    and for broadening $\Gamma/\pi=52$~meV. 
    (B) Spectral function at low energies for different values of the broadening $\Gamma$ and for 
    $\delta\varepsilon_{\rm C}=0.2$~eV.
    (C) Derivatives of spectral functions shown in (B) close to the inelastic spin excitation step.
     The dashed grey line marks the position of the bare excitation energy.
    (D) Shift and width of the ISTS step feature as a function of the broadening $\Gamma$ for $\delta\varepsilon_{\rm C}=0.2$~eV.
    The dashed lines show power-law fits to the data which yield exponents of $\alpha\sim0.7$ for both shift and width as a function
    of $\Gamma$.
    (E) Local spectral functions $\rho_i(\omega)$ for the two carbon sites of the NG shown in the inset
    for $\Gamma/\pi=39$~meV and $\delta\varepsilon_{\rm C}=0.2$~eV.
    (F) Density map of spatially resolved spectral function $\rho({\Br};\omega_-)$ evaluated
    at $z=5$\r{A} above the molecular plane for $\omega_-=-15.5$~meV. Other parameters as in (E).
  }
\end{figure*}

First, we consider a NG consisting of two coupled triangulenes as shown schematically in Fig.~\ref{fig:structures}(A).
Such triangulene dimers deposited on Au substrates were recently studied experimentally by STM spectroscopy, and theoretically by configuration 
interaction of an isolated molecule described by the Hubbard model (\ref{eq:NG})~\cite{mishra2020}.
Here we investigate the effect of the substrate on the $dI/dV$ and $d^2I/dV^2$ spectra, and in particular
the renormalization of the spin excitation energies due to the coupling to the substrate. 

The previous study showed that each of the triangulene units of the dimer carries a spin-1 located in its two ZMs
due to strong intra-triangulene Hund's rule coupling $J_{\rm H}$~\cite{Ortiz:NL:2019}. On the other hand, a Coulomb-driven kinetic 
exchange mechanism involving the virtual occupation (de-occupation) of the empty (filled) molecular orbitals closest 
in energy to the ZMs gives rise to an effective antiferromagnetic coupling $J_{\rm eff}{\ll}J_{\rm H}$ between the 
spins $S=1$ of the individual triangulenes. $J_{\rm eff}$ quenches the total spin of the dimer, leading to an $S=0$ 
singlet GS, separated in energy by $J_{\rm eff}$ from an excited $S=1$ triplet state, and by $3J_{\rm eff}$ from 
an $S=2$ quintuplet state~\cite{mishra2020}.

Here we only take into account the four ZMs of the two triangulene units
and model the effective antiferromagnetic exchange interaction between the two triangulenes by a Heisenberg 
exchange interaction $\hat{H}_S = J_{\rm eff}\,\mathbf{S}_1\cdot\mathbf{S}_2$. We take the value for $J_{\rm eff}$
from the energy gap between the $S=0$ GS and the excited $S=1$ state of a configuration interaction calculation 
including the two molecular orbitals of the dimer closest in energy to the zero modes. Assuming a Hubbard interaction 
of $U=1.9|t|=5.13$~eV results in $J_{\rm eff}=18$~meV.

Fig.~\ref{fig:dimer} shows results for the dimer model calculated within OCA for different coupling strengths and
on-site energies $\varepsilon_{\rm C}$ at low temperature ($kT=0.1{\rm meV}\sim1.2$K). Panel A shows 
the effect of detuning, $\delta\varepsilon_{\rm C}=\varepsilon_{\rm C}-\varepsilon_{\rm C}^\ast$, of the on-site energy $\varepsilon_{\rm C}$ away from 
the particle-hole (ph) symmetric point, $\varepsilon_{\rm C}^\ast=-485$~meV, on the spectral function on a large energy 
scale, including the Coulomb peaks. As the detuning $\delta\varepsilon_{\rm C}$ increases, the spectra become more asymmetric:
the lower Coulomb peak (corresponding to the electron removal energy) moves closer to the Fermi level, while the 
upper peak (electron addition energy) moves farther away from it.
As expected the value of the Coulomb gap of $\sim1.4$~eV in our model slightly underestimates the experimentally 
measured value of about $\sim1.65$~eV due to the neglect of the long-range part of the Coulomb interaction. 
A phenomenological way to fix the Coulomb gap was given in Ref.~\onlinecite{Ortiz:PSS:2020}.

We now focus on the low-energy part of the spectra, specifically on the lowest spin excitation. 
We choose a detuning of $\delta\varepsilon_{\rm C}=0.2$~eV which brings the lower Coulomb peak approximately 
into the same position as the experimentally observed one at about $-0.4$~eV. Fig.~\ref{fig:dimer}B
shows the spectral function for this choice of $\varepsilon_{\rm C}$ and for different values of the broadening 
by the substrate $\Gamma$. The spectral functions show two steps (one at negative and one at positive 
energy) similar to those found in the experiment~\cite{mishra2020}. These steps correspond to inelastic 
spin excitations from the $S=0$ singlet GS to the excited $S=1$ triplet state. 
 As the single-particle broadening $\Gamma$ increases, the inelastic steps 
 move inwards to lower energies, due to the renormalization induced by the coupling to the substrate,
 departing from the bare excitation energy of $\sim18$~meV,  and their triangular shape associated with
 Kondo exchange coupling becomes more pronounced. 

The renormalization of the excitation energies due to the coupling to the bath can be 
better appreciated in the derivative of the spectral function (Fig.~\ref{fig:dimer}C) which shows 
a peak at the excitation energy. Clearly, the peak moves to lower energies and broadens, as the coupling
to the bath $\Gamma$ increases. Fig.~\ref{fig:dimer}D shows the shift of the excitation energy and 
the broadening of the step extracted from the derivative of the spectral function (Fig.~\ref{fig:dimer}C)
as a function of $\Gamma$. Interestingly, we find a power-law behavior $\sim\Gamma^{0.7}$ for both shift and 
width of the step. At first sight this seems to contradict 2nd order perturbation theory results for the 
similar case of renormalization of the single-ion magnetic anisotropy of single magnetic adatoms on conducting substrates
which predict a quadratic behavior~\cite{Oberg:NNano:2014,Delgado:SS:2014}.
However, the perturbation theory results are expected to be valid only in the weak coupling (small $\Gamma$) regime to which we do not have 
access due to numerical issues~\footnote{For smaller $\Gamma$ we need increasingly fine frequency meshes to properly 
resolve the low-energy PP peaks which leads to higher computational cost as well as to numerical instabilities in the solution of the OCA equations.}.
It is therefore conceivable that the power-law behavior turns into quadratic behavior for very small $\Gamma$.

Importantly, our results show that the bare spin excitation energies are always renormalized due to the coupling
of the NG to the conduction electron bath in the substrate. The magnitude of the renormalization is, according to
our calculations, larger than the broadening of the spectral function peak. This is similar to the renormalization of the 
single-ion magnetic anisotropy measured in STM spectroscopy of single magnetic atoms and molecules on 
conducting substrates~\cite{Oberg:NNano:2014}.
In this sense the excitation energy measured in an STM experiment is never really the intrinsic one of the NG
since coupling to the conduction electrons in the substrate cannot be avoided in an STM setup. 
Introducing insulating layers between the NG and the conducting substrate in order to decouple the NG from 
the conduction electrons can significantly  reduce this effect, and thus allow to measure the bare excitation
energies to a better approximation. 

We now investigate the spatial dependence of the spectra. Since the triangulene ZMs are mainly localized
at the edge atoms of the NGs (cf. Fig.~\ref{fig:structures}), we expect that the spectral signature of
the inelastic spin excitations is localized there. This is 
indeed the case as can be seen in Fig.~\ref{fig:dimer}E which shows the local spectral density 
$\rho_i(\omega)$ for two carbon sites, one at the edge and one in the interior of the NG. Both show inelastic 
spin excitation steps, but the amplitude for the border site is significantly larger than for the one in 
the interior by almost one order of magnitude. Fig.~\ref{fig:dimer}F shows a density map of the 
spatially resolved spectral function $\rho({\bf{r}};\omega)$ evaluated at the energy of the maximum of the left step, 
$\omega_-=15.$~meV. Clearly, the density map resembles the spatial distribution of the ZMs (cf. Fig.~\ref{fig:structures}).

\subsection{Triangulene monomer}

\begin{figure}[t]
  \includegraphics[width=\linewidth]{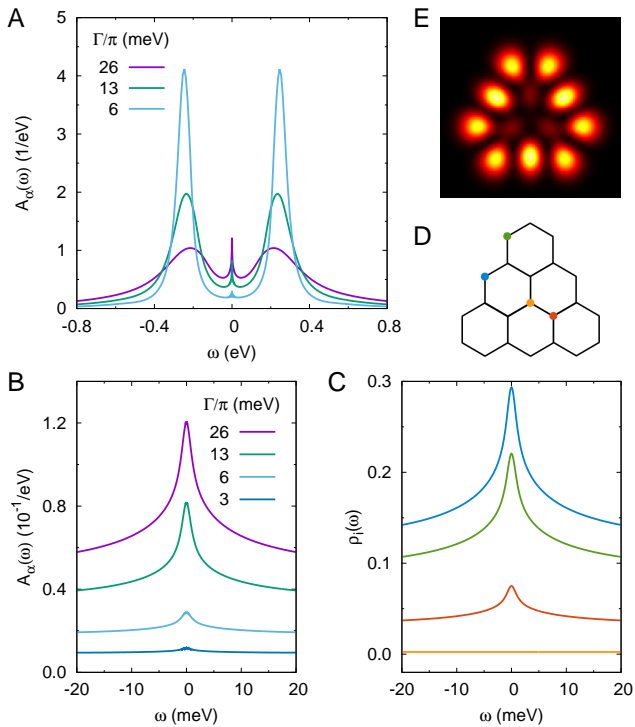}
  \caption{\label{fig:monomer}
    Spectral functions calculated within OCA for the triangulene monomer for $U=|t|$ and $t=-2.7$~eV at ph symmetry ($\varepsilon_{\rm C}=-0.26$~eV).
    (A) Spectral function of ZMs for different values of the coupling to the substrate $\Gamma$
    at low temperature $kT=0.4$~meV$\sim5$K.
    (B) Same as (A) but for a smaller energy scale showing the evolution of the Kondo peak as $\Gamma$ decreases.
    (C) Local spectral function for the carbon sites marked in the atomic structure (D) in the corresponding color.
    (E) Density map of spatially resolved spectral function $\rho({\Br};\omega_0)$ evaluated
    at $z=5$\r{A} above the molecular plane for $\omega_0=0$.
  }
\end{figure}

Next we consider a single triangulene molecule, shown in Fig.~\ref{fig:structures}C,
i.e., the basic building block of the triangulene dimer considered in the previous section.
Since now the antiferromagnetic exchange between the triangulene units is absent, the spin 
$S=1$ carried by the two ZMs of the triangulene remains unquenched.  Our results show a Kondo peak.
We note that Kondo peak measurements have been reported in at least two
spin-1 NGs on metallic substrates~\cite{li2020,heptauthrenesynth}, but not on triangulenes.
However, so far STM spectra of single triangulene molecules have only been 
measured for non-metallic Xe(111) as substrate, and no Kondo effect was found in this case. 
This is to be expected as the coupling to the sample electrode via the 'insulating' Xe layers 
or to the tip electrode via the vacuum is very weak. Due to exponential decay of the Kondo 
temperature with the ratio $\delta{U}/\Gamma$ (where $\delta{U}$ is the charging energy),
$T_K\sim{e^{-\delta{U}/\Gamma}}$, it becomes vanishingly small for $\Gamma\ll{\delta{U}}$.
Moreover, according to Nevidomskyy and Coleman for systems with $S>1/2$ the Kondo temperature
is exponentially reduced compared to a spin-1/2 Kondo system~\cite{Nevidomskyy:PRL:2009}. 

Nevertheless, if the coupling to the conduction electrons becomes sufficiently strong, for example
for a triangulene on a metallic substrate, the Kondo effect may occur at experimentally accessible
temperatures. Also, on a metallic substrate the Coulomb interaction $U$ will be reduced due to 
screening by the conduction electrons, thus further enhancing the Kondo temperature. 
In Fig.~\ref{fig:monomer}B we show the effect of changing the coupling to the substrate $\Gamma$ 
while keeping the Coulomb interaction fixed to $U=|t|$.
Panel A shows the spectral function of the ZMs for different values of $\Gamma$ at low 
temperature $T\sim5$K: The spectra show lower and upper Coulomb peaks at $\omega\sim\pm300$~meV and
a Kondo peak at $\omega=0$. As $\Gamma$ is increased the Coulomb peaks broaden and the 
Kondo peak becomes more pronounced. The latter can be better appreciated in panel B which
shows the evolution of the Kondo peak as $\Gamma$ changes. For the largest coupling
shown in Fig.~\ref{fig:monomer} ($\Gamma/\pi=26$~meV) we obtain a Kondo temperatures
estimated from the halfwidth of the Kondo peak of about $T_K\sim15$K,
well above typcial temperatures that can be achieved in STM experiments.
Also note that the critical magnetic field for observing the Zeeman splitting of the
Kondo peak in the $S=1$ Kondo case is given by $B_c\sim k\,T_K/2\mu_B$. Thus for Kondo temperatures
of up to 15K obtained here, fields of up to 11 Tesla are required to observe the splitting experimentally.
For the lowest $\Gamma$ considered here, 
the Kondo peak vanishes at the chosen temperature $T\sim5$K. We note that inelastic steps,
corresponding to transitions from the S=1 to S=0, with energies in the range of the intramolecular
exchange of about 260meV, cf. Fig.~6c,d in Ref.~\onlinecite{Ortiz:PSS:2020},
are not visible in the spectra. They are obscured by the much more pronounced
Coulomb peaks which happen to reside just at the same energy.

Fig.~\ref{fig:monomer}C shows the local spectral functions $\rho_i(\omega)$ projected on
different carbon sites $i$ marked by circles in corresponding color in the atomic structure 
in panel D. In correspondence to the weight of the site in the ZM wave function
$\psi_\alpha(i)$, cf. Fig~\ref{fig:structures}C, the Kondo peak becomes less and less 
pronounced as we move from the blue, to the green, and to the red site. On the other hand,
the Kondo peak is completely absent on the orange site which belongs to a different sublattice
than the ZMs. 
The density map in Fig.~\ref{fig:monomer}E shows the spatially resolved spectral function evaluated 
at $\omega=0$. As the Kondo peak only shows up at the sites contributing to the ZMs, the
map reflects the density associated with the wavefunctions of the ZMs, c.f. Fig~\ref{fig:structures}C.

\subsection{'Rocket' structure}

\begin{figure}[t]
  \includegraphics[width=\linewidth]{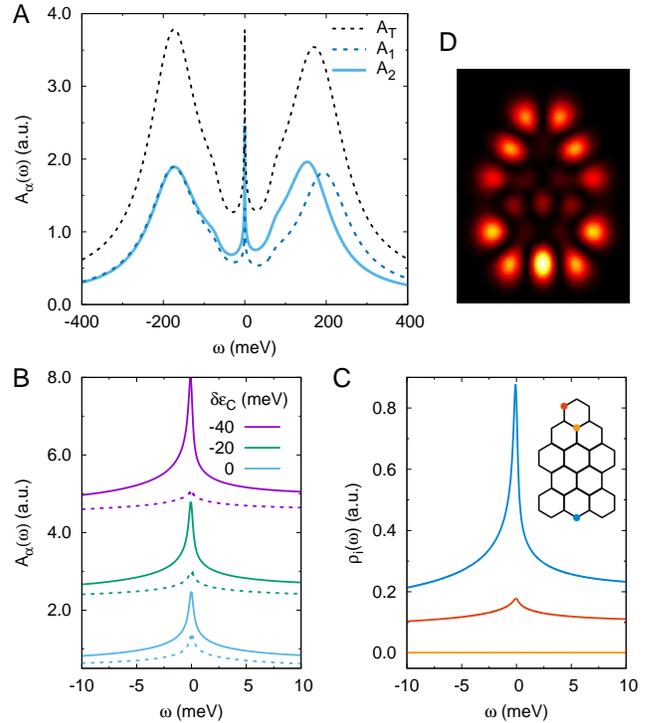}
  \caption{\label{fig:rocket}
    Spectral functions calculated within OCA for rocket structure for $U=|t|$ and $\Gamma/\pi=13$~meV
    at low temperature, $kT=0.1$~meV$\sim1.2$K.
    (A) Orbital resolved ($A_1$,$A_2$) and total ($A_{\rm T}=A_1+A_2$) spectral functions
    at half-filling ($N=N_1+N_2=2$) at $\varepsilon_{\rm C}=-0.186$~eV.
    (B) Orbital resolved spectral functions around the Kondo peak for different values of detuning $\delta\varepsilon_{\rm C}$ from
    half-filling. The dashed (solid) lines show the spectral functions of orbital 1 and 2, respectively.
    (C) Local spectral function $\rho_i(\omega)$ for carbon sites $i$ marked in corresponding color in the atomic structure 
    shown in the inset for $\delta\varepsilon_{\rm C}=-40$~meV.
    (D) Density map of spatially resolved spectral function $\rho({\Br};\omega_0)$ evaluated at $z=5$\r{A}
    above the molecular plane for $\omega_0=0$.
  }
\end{figure}

Finally, we consider the 'rocket' structure shown in Fig.~\ref{fig:structures}E, which has recently been studied
experimentally by STM spectroscopy~\cite{li2020}. As reported there, we also find a spin-1 Kondo effect signaled
by a sharp Abrikosov-Suhl resonance at the Fermi level at low temperatures, see Fig.~\ref{fig:rocket}(A).
Similar to triangulene discussed in the previous section, the 'rocket' structure hosts a spin-1 in its two ZMs $\psi_1$ and $\psi_2$.
However, different from triangulene, $\psi_1$ and $\psi_2$ have very different densities with $\psi_1$
mostly localized in the upper part and $\psi_2$ mostly localized in the lower part of the NG, as can be seen in
Fig.~\ref{fig:structures}E. Interestingly, $\psi_2$ is also somewhat more delocalized than $\psi_1$. This leads to a slightly
smaller intra-orbital Coulomb repulsion for the former ($U_2\sim300$~meV) than for the latter ($U_1\sim350$~meV), as these matrix
elements are inversely proportional to extension of the orbitals~\cite{Ortiz:NL:2019}.
Accordingly, the upper Coulomb peaks of both orbitals corresponding to the addition of one electron over the half-filled
($N=2$) GS are split in energy by $\sim{U_1-U_2}\sim50$~meV, as can be seen in Fig.~\ref{fig:rocket}(A). On the other hand,
the lower Coulomb peak is at the same position for both orbitals due to the degeneracy of their energy levels.

As the addition energies $\delta{U}_\alpha$ differ for both orbitals, so do their respective Kondo temperatures,
according to $T_{K,\alpha}\sim e^{-\delta{U}_\alpha/\Gamma}$. Taking into account that the broadening $\Gamma$ is equal
for both orbitals, we thus have $T_{K,1}<T_{K,2}$. Indeed the Kondo peak for $\psi_1$ is considerably less pronounced
than the one for $\psi_2$, as can be seen in Fig.~\ref{fig:rocket}(B). Moreover, moderately lowering the energy levels
of both orbitals, the disparity of both Kondo peaks is further enhanced for the following reason: At half-filling $N\sim2$,
the individual occupations of both orbitals are slightly different due to their different Coulomb interactions,
namely $\psi_1$ is slightly less than half-filled, $N_1\sim0.99$, while $\psi_2$ is slighlty more than half-filled, $N_2\sim1.01$.
Upon lowering the energy of both orbitals, $N_1$ is at first driven closer to half-filling, while $N_2$ is driven further
away from half-filling. Additionally, the higher charging energy of $\psi_1$ leads to $N_1$ increasing slower than
$N_2$ upon lowering of the energy levels. Thus charge fluctuations and correspondingly the Kondo temperature are reduced for orbital $\psi_1$,
while they are  enhanced for orbital $\psi_2$. For temperatures $T_{K,1}<T<T_{K,2}$ this leads to \emph{partial} Kondo
screening where only the spin-1/2 in $\psi_2$ is screened. For the largest detuning considered here ($\delta_{\rm C}=-40$~meV)
we estimate $T_{K,2}\sim7.5$K from the halfwidth of the Kondo peak of orbital $\psi_2$ in very good agreement with the
experimental value obtained by Li and coworkers~\cite{li2020}.

As the two ZMs are localized in different parts of the NG, the different manifestion of the Kondo peaks in 
the two orbitals leads to strongly site dependent spectral functions, as can be seen in Fig.~\ref{fig:rocket}(C):
While sites in the lower part of the NG corresponding to $\psi_2$ (e.g. blue site) show a very pronounced Kondo peak,
the Kondo peak is strongly suppressed for sites in the upper part (e.g. red site) of the NG corresponding to $\psi_1$.
On the other hand, sites that live on a different sublattice than the ZMs (e.g. orange site) do not show a Kondo peak at all.
The density map of the spatially resolved spectral function $\rho(\Br;\omega_0=0)$ summarizes
these findings: due to the predominance of the Kondo peak in the lower part of the NG the density map is considerably
brighter in the lower part than in the upper part. On the other hand, the inner part of the NG remains mostly dark
as the ZMs predominantly live at the border of the NG.

\section{Conclusions}

We have studied the effect of the coupling to a metallic substrate on the spectra of open-shell NGs.
For NGs with a non-degenerate $S=0$ GS such as the triangulene dimer, the exchange coupling to the conduction electrons
in the substrate leads to the reduction  of the energies of the  $S=0\rightarrow S=1$ excitations, compared with the bare
energies computed ignoring the coupling to the substrate. On the other hand,
for NGs with a degenerate GS, coupling to the conduction electrons leads to Kondo effect, fully or partially screening
the spin of the NG, if the coupling is strong enough and/or the temperature is low enough. In very symmetric
situations such as in the triangulene molecule the Kondo screening is equally strong for all ZMs, generally
leading to a situation where the entire spin $S=1$ of the NG is screened by the Kondo effect. In this case the Kondo
temperature is exponentially reduced compared to the spin-1/2 Kondo effect~\cite{Nevidomskyy:PRL:2009},
thus requiring very low temperatures and/or very strong coupling to the substrate to become measurable.
For less symmetric situations such as the 'rocket' structure, the Kondo screening strength may vary
considerably among the ZMs due to their different intra-orbital Coulomb interactions, leading to partial Kondo
screening of the NG spin as reported in recent experimental work~\cite{li2020}. Our results further indicate that in this
case the partial Kondo screening situation leads to manifestation of the Kondo peak predominantly at carbon sites in the
lower part of the NG associated with the Kondo screened ZM. 

Our results stress the need to consider the coupling to the substrate in order to account for IETS experimental results, and anticipate
discrepancies with high level quantum chemistry calculations for gas phase molecules. Our formalism also predicts the broadening
of the spectral features that ultimately reflects the finite lifetime of the spin excitations due to Kondo exchange with the
substrate and will have to be considered in proposals to use nanographenes for quantum information
purposes~\cite{lombardi2019,gaita2019}, whenever nanographenes are to be contacted to electrodes.

\begin{acknowledgments}
  DJ acknowledges funding by the Basque Government through grant ``Grupos Consolidados UPV/EHU del Gobierno Vasco'' (Grant No. IT1249-19).
  RO acknowldedges financial support from Generalitat Valenciana and Fondo Social Europeo (Grant No. ACIF/2018/175). 
  JFR acknowledges financial support  from Generalitat Valenciana (Prometeo2017/139),
  and MINECO-Spain (Grant No. PID2019-109539GB-C41).
  JFR and DJ acknowledge  funding from  FEDER/Junta de Andaluc\'{i}a-Consejer\'{i}a de
  Transformaci\'{o}n Econ\'{o}mica, Industria, Conocimiento y Universidades, grant P18-FR-4834.
  We acknowledge fruitful discussions with G. Catarina and J.C. Sancho-Garc\'{i}a.
\end{acknowledgments}

\appendix

\section{One-crossing approximation}
\label{app:OCA}

We solve the multi-orbital AIM within the one-crossing approximation (OCA)~\cite{Haule:PRB:2001,Haule:PRB:2010}.
The first step is a numerical diagonalization of the \emph{isolated} impurity Hamiltonian (\ref{eq:HC}),
${\cal H}_{\rm C}\ket{m}=E_m\ket{m}$ for different fillings $N=\sum_\alpha \langle N_\alpha \rangle$ of the impurity shell. 
Here we consider the NG close to half-filling. 
The coupling to the substrate ${\cal V}_{\rm hyb}$ only connects eigenstates with occupations differing 
by one electron, leading to charge and spin fluctuations in the impurity shell.
Thus for a NG with $M$ ZMs we consider the occupations $N=M$ and $N=M\pm1$.
It is the fluctuations between the impurity GS manifold and excited states with one more or
one less electron that give rise to both Kondo effect and renormalization of the spin excitation energies.

In the next step a diagrammatic expansion of the many-body eigenstates $\ket{m}$  of the (isolated) impurity 
${\cal H}_{\rm C}$ in terms of the hybridization with the substrate is developed. 
To this end one introduces so-called pseudo-particles (PPs) $m$ corresponding to
the many-body eigenstates $\ket{m}$. The full propagator of such a PP $m$ can be
written as $G_m(\omega) = 1/(\omega-\lambda-E_m-\Sigma_m(\omega))$
where $\Sigma_m(\omega)$ is the PP self-energy which describes the
renormalization (real part) and broadening (imaginary part) of the
PP $m$ due to the interaction with other PPs $m^\prime$ mediated by
the conduction electron bath.
$-\lambda$ is the chemical potential for the PPs which has to be adjusted
such that the total PP charge $Q=\sum_ma_m^\dagger a_m$ is conserved,
imposing the completeness of the many-body Hilbert space.

OCA consists in a diagrammatic expansion of the PP self-energies $\Sigma_m$
in terms of the hybridization function $\Delta_\alpha(\omega)$ to
infinite order but summing only a subset of diagrams (only those involving
conduction electron lines crossing at most once).
This leads to a set of coupled integral equations for the PP propagators
and self-energies that have to be solved self-consistently.

Once the OCA equations are solved, the real electron spectral function
$A_\alpha(\omega)$ for the NG ZMs are obtained from convolutions of PP propagators
$G_m(\omega)$.
The PP resonances for the GS manifold feature sharp resonances at the renormalized many-body 
energies $E_m^\ast=E_m+\Re\,\Sigma_m(E_m^\ast)$. 
The differences between the renormalized energies of the excited states $E_m^\ast$
and the GS $E_0^\ast$ yield the actual electronic excitations of the system. 
Thus we see that the coupling to the conduction electron bath leads to the renormalization
of the spin excitations, described by the real part of the PP self-energies
within our approach. More details on the application of the OCA method
to nanoscale quantum magnets can be found e.g. in Refs.~\onlinecite{Jacob:EPJB:2016,Jacob:PRB:2018}.

\bibliographystyle{apsrev4-2}
\bibliography{biblio}{}

\end{document}